\documentclass[aps,showpacs,twocolumn,prd,superscriptaddress]{revtex4}

\usepackage{amsmath}
\usepackage{latexsym}
\usepackage{graphicx}

\begin{document}

\title{Comparisons of binary black hole merger waveforms}

\author{John G. Baker}
\affiliation{Gravitational Astrophysics Laboratory, NASA Goddard Space Flight Center, 
8800 Greenbelt Rd., Greenbelt, MD 20771, USA}

\author{Manuela Campanelli}  
\affiliation{Center for Gravitational Wave Astronomy, 
Department of Physics and Astronomy,
The University of Texas at Brownsville, Brownsville, Texas 78520}
\affiliation{Center for Computational Relativity and Gravitation,
School of Mathematical Sciences,
Rochester Institute of Technology, 78 Lomb Memorial Drive, Rochester,
 New York 14623}

\author{Frans Pretorius}
\affiliation{Department of Physics, University of Alberta, Edmonton, AB T6G 2G7 Canada}
\affiliation{Canadian Institute for Advanced Research, Cosmology and Gravity Program}
\affiliation{Department of Physics, Princeton University, Princeton, NJ 08540} 

\author{Yosef Zlochower}
\affiliation{Center for Gravitational Wave Astronomy, 
Department of Physics and Astronomy,
The University of Texas at Brownsville, Brownsville, Texas 78520}

\date{\today}

\begin{abstract}
This a particularly exciting time for gravitational wave physics.
Ground-based gravitational wave detectors are now operating at a
sensitivity such that gravitational radiation may soon be directly
detected, and recently several groups have independently made
significant breakthroughs that have finally enabled numerical
relativists to solve the Einstein field equations for
coalescing black-hole binaries, a key source of gravitational
radiation. The numerical relativity community is now in the position
to begin providing simulated merger waveforms for use by the 
data analysis community,
and it is therefore very important that we provide ways to validate the
results produced by various numerical approaches. Here, we present a
simple comparison of the waveforms produced by two very
different, but equally successful approaches---the generalized harmonic
gauge and the moving puncture methods. We compare
waveforms of equal-mass black hole mergers
with minimal or vanishing spins. The results show exceptional
agreement for the final burst of radiation, with some differences
attributable to small spins on the black holes in one case.
\end{abstract}

\pacs{EDIT
04.25.Dm, 
04.30.Db, 
04.70.Bw, 
95.30.Sf, 
97.60.Lf  
}

\maketitle

The intense gravitational radiation produced by merging
black-hole-binaries is expected to be among the strongest
gravitational waves produced by astrophysical systems, and may
even be detectable by the current generation of gravitational wave
observatories, such as initial LIGO~\cite{LIGO} (which is already
taking data at its designed sensitivity). However, any detection with
the current instruments is likely to be at a low signal-to-noise
ratio, and therefore having accurate information about the
predicted waveforms will increase the observers' ability to discern
the signals from noise, effectively increasing the likelihood of an
identifiable detection. For the more sensitive instruments to come,
including advanced LIGO, and the space-based Laser Interferometric
Space Antenna (LISA) observatory~\cite{Danzmann:2003tv}, better
knowledge of the waveform predictions from General Relativity will
increase the precision with which source parameters (i.e.\ the masses
and spins of the components and the eccentricity of the their orbit)
can be identified from the waveform, and will make high-precision
tests of General Relativity in the strong-field regime possible.

Recently several groups have independently made significant
breakthroughs~\cite{Pretorius:2005gq,Campanelli:2005dd,Baker:2005vv}
in solving the Einstein field equations for coalescing black-hole
binaries. Each of these groups were able to simulate the entire merger
phase for various black-hole-binary configurations and accurately
calculate the resulting waveforms.  The novelty of the new techniques,
however, leaves some room for questions of whether the results might
be influenced by modeling errors that were not evident in the context
of an individual research group's effort.  In this paper we compare
the recently published results of three groups, which have each
independently solved the problem of how to successfully evolve
black-hole-binary systems through the last orbits, merger, and ringdown to
extract the gravitational radiation.  Each of these groups has
performed simulations of systems of equal-mass black hole
mergers~\cite{Campanelli:2006gf, Baker:2006yw, Buonanno:2006,
Scheel:2006gg, Bruegmann:2006at} with the UTB and GSFC groups studying non-spinning
irrotational binaries, while Pretorius applied a corotational binary
model, implying a small spin $a/m=0.08$ on each of the individual black holes.
All simulations went through between 1.75 and 2.5
orbits before merger. Each of the groups has gone to some effort to
demonstrate the quality of their waveforms, giving at least a rough
estimate of the level of error in their simulations, but comparing the
results gives us a more comprehensive understanding of the full range
of modeling uncertainties.

If we look at waveform predictions as a modeling process, it is clear
that there are several different classes of potential errors
roughly associated with different steps in the modeling process.  A
first step is to pose initial data describing the configuration of the
black holes at some point a few orbits before merger.  In our
comparisons, no attempt is made to fix the starting point of the
simulations. Rather, we will look for agreement among the runs for the
portion of the merger covered by all simulations.  The runs shown here
begin with a variety of initial data models. Pretorius uses the
Cook-Pfeiffer~\cite{Caudill:2006hw,Pfeiffer:2002wt} technique to
generate quasi-circular data using the conformal-thin-sandwich
decomposition with excised horizon interiors. The GSFC and UTB
groups use the Brandt-Br\"ugmann puncture approach~\cite{Brandt97b} to
generate the initial data (which, in turn, is based on the
Bowen-York~\cite{Bowen80} ansatz for the extrinsic curvature). The
GSFC and UTB approaches are distinguished by their respective
choices of normalization and the methods used to determine the free
initial data parameters. The GSFC group obtained values for the
puncture positions and momenta from the Cook initial data
sequence~\cite{Cook94}. The UTB group, on the other hand, used the 3PN
trajectories of quasi-circular binaries with an orbital period of $M
\omega = 0.050$ and total mass $M_{ADM} =1$ to determine the puncture
positions, mass parameters, and momenta. The two choices of parameters
turn out to be very similar; differing by less than 1\% after
re-normalizing to unit mass (the overall scaling is not important here
because the waveforms will be rescaled by the final mass of the
remnant). 

The question here is how sensitive the waveform results are to
variations in the initial data model.  While it has not been straight
forward for the groups individually to compare puncture models with
Cook-Pfeiffer data, it is possible for each group to study initial data
sensitivity within the class of initial data they are using.  This has
been addressed by the GSFC group by performing a range of
simulations varying from the one presented here, to a four orbit run.
Their simulations showed excellent agreement at roughly a 1\% level
beginning about $50M$ before the peak of merger radiation and onward.
Prior to this late merger part of the waveforms, their waveforms showed rough
agreement with variations that seemed to be associated with eccentric
motion in the binaries that depended on the specifics of the initial
data parameters for each run. Similar estimates of the intrinsic
differences in Cook-Pfeiffer initial data sets were presented 
in~\cite{Buonanno:2006} by evolving three different initial separations.
Good agreement was also found near the peak of energy emission, though
given that the corotation condition gave the initial black holes spins
of $0.06$, $0.08$ (the simulation examined here) and $0.11$, the agreement
is not expected to be as close as the GSFC results. 

All three approaches used conformally flat initial data. These data contain
a short burst of spurious radiation that quickly leaves the system.
The amplitude of this spurious signal decreases monotonically as the
initial binary separation is increased and does not appear to alter
the binaries waveform in a significant way after the pulse is radiated
away.

The next, crucial stage in the modeling is to perform a numerical simulation
of the merger dynamics. Pretorius' evolutions were performed
using a generalized harmonic evolution scheme~\cite{Pretorius:2004}
with constraint damping~\cite{gundlach:2006}.
Both the UTB and GSFC groups use the BSSN~\cite{Nakamura87,
Shibata95, Baumgarte99} system of equations in conjunction with a
$1+\log$ type lapse and $\Gamma$-driver shift conditions. The UTB
method differs from the GSFC method in that the standard BSSN
conformal exponent $\phi$ (which has an $O(\log r)$ singularity at the
punctures) is replaced by the initially $C^4$ field $\chi =
\exp(-4\phi)$, and by a different choice for the precise form of the
gauge conditions.  The GSFC and Pretorius simulations took advantage of Adaptive 
Mesh Refinement, to allow a distant outer boundary without reducing 
resolution near the black holes, while the unigrid UTB simulations used a
`multiple transition' fisheye transformation~\cite{Campanelli:2006gf}
to push the physical boundaries to $216M_{ADM}$ (simulating FMR).

The final modeling step is to determine gauge invariant waveform
information from the simulation data that represents the gravitational
wave signal seen by distant observers.  This is complicated by the
fact that typical simulations require the radiation information be
measured from the simulation data at relatively small distances from
the sources in comparison to the far larger distances between
earth-based observers and astrophysically realistic merging binaries.
Again our groups have taken different approaches to this task.
Working alone a group can quantify error levels by comparing waveforms
extracted at different distances from the sources and by comparing the
radiated energy and angular momentum with the mass and angular
momentum differences between the remnant black hole mass and angular
momentum and
the ADM mass and angular momentum. In particular, the deviation
between this mass difference and the radiated energy is a measure of
the amplitude error in the waveform, while the deviation between this
angular momentum difference and the radiated angular momentum is a
measure of the phase error in the waveform as well as the amplitude
error.

The UTB group used both the coordinate-based~\cite{Baker:2001sf} and the
quasi-Kinnersley~\cite{Beetle:2004wu,Campanelli:2005ia} tetrads to
define $\psi_4$ (both extracted at $r=30M$), and found no significant
difference in the dominant $(\ell=2, m=\pm2)$ modes for the two choices of
tetrads. The coordinate-based tetrad is equivalent to the Kinnersley tetrad
for Schwarzschild spacetimes, but the resulting Newman-Penrose
$\psi_4$ contains contributions from the Kinnersley $\psi_0$ ---
$\psi_3$ for spacetimes with spin. The quasi-Kinnersley tetrad removes
this type of mixing but leaves overall phase and amplitude ambiguities
in $\psi_4$. However, the good agreement between the radiated angular
momentum and mass and the angular momentum and mass losses of the
remnant black hole indicate that these amplitude and phase ambiguities
are small for the $(\ell=2,m=\pm2)$ modes.
They verified the accuracy of the waveforms that they produced by
evolving the binary with three resolutions and measuring the deviation
in the waveforms between resolutions. They also confirmed that the
radiated mass $(3.5\pm0.1)\%$ and angular momentum $(27\pm1)\%$
calculated from the waveform in their simulations matched the
mass $(3.5\pm0.1)\%$ and angular momentum $(26.9\pm0.3)\%$ losses of the
 remnant horizon. The uncertainties in the radiated mass and angular
momentum arise from extracting these quantities at finite distances
and extrapolating to infinity.
The uncertainty in the waveform due to finite resolution is quite small.
After translating the waveforms so that the maximum amplitudes coincide
and correcting for the phase differences at this maximum,
they find that the plunge part of the waveforms differ by $\sim0.25\%$
between resolutions of $M/22.5$ and $M/27$, while the late-time
ringdown part of the waveforms differ by $\sim 1.4\%$. The radiated 
energy and radiated angular momenta differ by $0.7\%$ and $0.2\%$
respectively.

The GSFC waveform simulations were carried out at three resolutions with 
finest-grid resolutions of $h_f\in\{3M/64,M/32,3M/128\}$.  At second-order 
convergence, the difference between these two higher resolution simulations
provides a estimate for the error of the higher-resolution run.  The peak 
wave amplitudes for these runs showed agreement to $\sim0.5\%$, with
differences of $\lesssim1\%$ in the waveforms overall after shifting 
to align the peaks. The GSFC waveforms were extracted using 
a coordinate based tetrad, as in~\cite{Baker:2001sf}, but extractions were
carried out on larger spheres $r/M\in\{20,30,40,50\}$.
Results from both higher-resolution runs including the three extraction radii 
with $r\ge30M$ indicate consistent results for the radiative losses 
of energy $(3.44\pm.01)\%$ and angular momenta $(26.0\pm.3)\%$.
In the comparisons we show waveforms from the $h_f=3M/128$ run 
extracted at $r=40M$, which strikes a good balance between $1/r$ effects
too close in, and numerically induced dissipation as the waves propagate 
farther out.

In the generalized harmonic code waveforms were also extracted
using a coordinate based tetrad at extraction radii 
from $r=13M$ through $r=51M$, as described in~\cite{Buonanno:2006}. 
Convergence tests at an extraction radius of $r=51M$ suggest
a maximum amplitude error of $9\%$, a maximum bare
phase error of $0.7$ cycles, a maximum phase error after time
translation and rotation
of $0.06$ cycles, $(4.3\pm0.4)\%$ total energy radiated and 
$(33\pm2.1)\%$ total angular momentum radiated.

We will compare the results of one run from each group.
Table~\ref{table:stats} summarizes the runs compared here.  As
stated, the runs agree in mass ratio and have vanishing or small
spins.  As we construct our comparison, we will initially ignore the
small difference in spins, though we will return, as we consider the
comparative results.  The other physical parameters, total mass,
merger-time, and merger-orientation are unrelated among our runs.
Fortunately, each simulation result can be viewed as representing an
equivalence classes of systems with different values for these
parameters.  Upon concretely simulating one member of the equivalence
class, we can derive the others by suitable application of
time-translation, rotation, and scaling operations.  Our comparisons
will focus directly on the strongly dominant spin-weighted spherical
harmonic component of the radiation, the $(\ell=2,m=2)$ component
(which is mirrored by the $(\ell=2, m=-2)$ component).  Following
Refs.~\cite{Baker:2002qf,Baker:2006yw} we will compare the
simulations by rescaling masses by the final remnant black hole mass, 
applying a
time-translation so that the time of peak waveform (polarization)
amplitudes agree (labeled $t=0$), and applying a rotation so that the
waveform phases agree at $t=0$. Note that a similar comparison method
was introduced in Ref.~\cite{Baker:2002qf}.

\begin{table}
\caption {Model Parameters for the three runs. $r_{\rm ext}$ is the
radius of the waveform extraction sphere, $r_{\rm bdry}$ is the location
of the outer boundary, $M_{ADM}$ is the total ADM mass of the
system, $J_{ADM}$ is the total ADM angular momentum, $S$ is the initial
horizon spins, $\tau$ is the initial
orbital period, and $\ell$ is the initial proper distance between 
the horizons.}
\begin{ruledtabular}
\begin{tabular}{llll}\label{table:stats}
 & Pretorius & GSFC & UTB \\
\hline
$\tau / M_{ADM}$ & 153 & 127 & $126$ \\
$\ell / M_{ADM}$ & 9.89 & $9.9$ & $10.0$ \\
$J_{ADM}/M_{ADM}^2$ & 0.933 & $0.873$ & $0.876$ \\
$S/M_{ADM}^2$ & 0.020 & 0 & 0 \\
$r_{\rm ext} / M_{ADM}$ & 50.6 & 40.2 & 30.0 \\
$r_{\rm bdry} / M_{ADM}$ &  $\infty$ & 771 & 216 \\
\end{tabular}
\end{ruledtabular}
\end{table}

\begin{figure}
\includegraphics[width=3.0in]{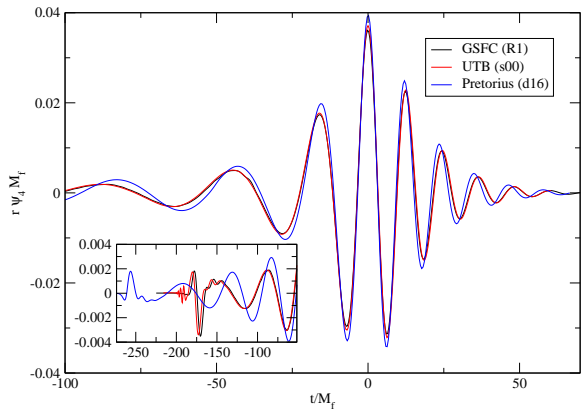}
\caption{The real part of the $(\ell=2, m=2)$ mode of the 
gravitational radiation waveforms from three different
research groups. The waveforms from the strictly non-spinning GSFC and
UTB runs show excellent agreement after $t=-50M_f$, while Pretorius'
run rings-down at slightly higher frequency consistent with the
corotating spins on the black holes. The inset shows the early part
of the waveform including the artificial pulse of radiation due to the
choices of initial data.}
\label{fig:waveform}
\end{figure}

\begin{figure}
\includegraphics[width=3.0in]{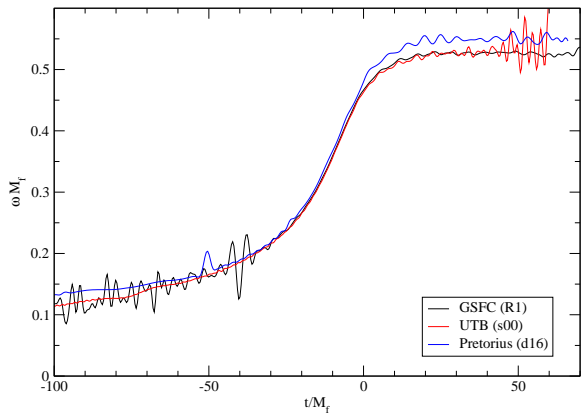}
\caption{Gravitational radiation polarization frequencies from 
three different research groups. 
The UTB and GSFC waveforms are very similar with the UTB result
showing high frequency oscillation at late times due to boundary
reflections. The consistently larger frequencies of Pretorius'
waveform are apparent.
}
\label{fig:frequency}
\end{figure}

In Fig.~\ref{fig:waveform} we show the real part of the $(\ell=2,m=2)$
mode of $\psi_4$ for the three different simulations after normalizing
the waveform by the final remnant mass, rescaling the $t$-axis by the final
remnant mass, translating the waveforms so that the maximum amplitude
is centered at $t=0$, and multiplying by constant phase factors. 
As expected the
UTB and GSFC waveforms are remarkably similar (i.e.\ they both start
from essentially the same configuration), while Pretorius'
waveform shows a higher frequency consistent with its larger spin.
Note that the initial data pulse evident in the inset is smaller for
Pretorius' waveform. This is mainly due to Pretorius' simulation
starting from further separated binaries. The UTB and GSFC runs
differ most strongly (but by less than $3\%$ of the amplitude) 
at the peak of the waveform. Small differences in the UTB and GSFC
waveform at the $O(1\%)$ level are expected because the initial data
were only in agreement to that order. In addition the GSFC and UTB
waveforms were extracted at different radii $r_{\rm ext}$,
and the extracted waveforms contain an $O(1/r_{\rm ext})$ error. 
Note that the plunge behavior ($-25 < t/M_f < 25$) is in large part
independent of differences in the starting configuration and initial
spins (at least for small spins). This independence of the plunge
waveform to small changes to the initial configuration is more pronounced
in Fig.~\ref{fig:frequency}, which shows the instantaneous frequency
of the $(\ell=2, m=2)$ mode of $\psi_4$ versus time. 
While Pretorius' simulation shows a generally consistent waveform,
especially during the plunge, the frequency is slightly higher than
that determined for the more similar GSFC and UTB waveforms during
both the initial orbital inspiral part of the waveform and the
late-time ringdown part of the waveform. We suspect that the
differences may primarily be the effect of spin on the black holes
in Pretorius' initial data, which tends to result in higher
frequencies in merger and ringdown~\cite{Baker:2003ds,
Campanelli:2006uy, Campanelli:2006fg}.  The
relatively large oscillations observed in the UTB frequency profile at
late times are due to boundary reflections, similar noise in the early
part of the GSFC waveforms are caused by echos of early gauge dynamics
off the AMR refinement interfaces.

From the above results we can make several important conclusions.
First, the plunge part of the waveform is a robust feature that varies
little with either the binaries starting configuration or the
component black hole spins (provided that the spins are small).
Such universality of the late part of the waveforms 
was shown in~\cite{Baker:2006yw} for a specific
sequence of initial configurations with increasing separation.  Our 
comparative results support the notion in a broader context.
Second, the spurious radiation content of the initial data does not
contaminate the merger waveform significantly.  Finally, the good
agreement between these waveforms supports each individual groups
claims of accuracy. 
On the other hand, our comparisons also seem to indicate 
that the GSFC and UTB waveforms agree more closely with each 
other than with Pretorius' results which realize slightly 
higher frequencies.  This difference is consistent with the small 
spins on the black holes in Pretorius' simulations. Note however
that the differences are in fact smaller than the conservative
estimates of the errors of the latter simulation presented 
in~\cite{Buonanno:2006}. Further comparisons
of runs with more closely-matched astrophysical parameters and higher
accuracy (both in terms of truncation error and controlling/understanding
systematics) will
be needed to verify that this is indeed the effect of astrophysical 
differences. Also, the comparisons presented here have been
rather qualitative, and more rigorous, quantitative procedures
will eventually be necessary---such a strategy was proposed 
in ~\cite{baumgarte:2006}, though the emphasis there was on
quantifying the accuracy and differences in waveforms with
respect to the data analysis effort. It will be interesting
to compare these waveforms, and those currently being
produced by other groups, with this methodology. However,
given that visually the three waveforms shown here are more similar
than the sample waveforms used in ~\cite{baumgarte:2006} we can
anticipate that the overlap of the waveforms in Fig.~\ref{fig:waveform}
will be greater than $0.95$ for black hole binaries with masses greater
than roughly $200$ solar masses (see Fig.'s 2 and 8 in ~\cite{baumgarte:2006}).

The type of waveform comparisons illustrated in this paper
will be facilitated by the NRwaves project~\cite{NRwaves},
which will contain a repository of waveforms produced by
all participating numerical relativity groups.

\acknowledgments
We thank Carlos Lousto and Pablo Laguna for careful reading of this text.
M.C. and Y. Z. gratefully acknowledge the support
of the NASA Center for Gravitational Wave Astronomy at University of
Texas at Brownsville (NAG5-13396) and the NSF for financial support
from grants PHY-0140326 and PHY-0354867. UTB simulations were
performed on the 70-node ``Funes'' cluster at UTB, on the ``Lonestar''
supercomputer at TACC, and on the ``Tungsten'' supercomputer at NCSA.
F.P's simulations were performed on
the University of British Columbia's ``vnp4'' cluster (supported by
CFI and BCKDF),  ``WestGrid'' machines (supported by CFI, ASRI and
BCKDF), and ``Lonestar''.
F.P. gratefully acknowledges support from the CIAR, NSERC
and Alberta Ingenuity.
\bibliographystyle{apsrev}
\bibliography{comparisons.v2.bib}

\begin{thebibliography}{29}
\expandafter\ifx\csname natexlab\endcsname\relax\def\natexlab#1{#1}\fi
\expandafter\ifx\csname bibnamefont\endcsname\relax
  \def\bibnamefont#1{#1}\fi
\expandafter\ifx\csname bibfnamefont\endcsname\relax
  \def\bibfnamefont#1{#1}\fi
\expandafter\ifx\csname citenamefont\endcsname\relax
  \def\citenamefont#1{#1}\fi
\expandafter\ifx\csname url\endcsname\relax
  \def\url#1{\texttt{#1}}\fi
\expandafter\ifx\csname urlprefix\endcsname\relax\def\urlprefix{URL }\fi
\providecommand{\bibinfo}[2]{#2}
\providecommand{\eprint}[2][]{\url{#2}}

\bibitem[{\citenamefont{Vogt}(1992)}]{LIGO}
\bibinfo{author}{\bibfnamefont{R.}~\bibnamefont{Vogt}}, in
  \emph{\bibinfo{booktitle}{Sixth {M}arcel {G}rossman Meeting on General
  Relativity (Proceedings, Kyoto, Japan, 1991)}}, edited by
  \bibinfo{editor}{\bibfnamefont{H.}~\bibnamefont{Sato}} \bibnamefont{and}
  \bibinfo{editor}{\bibfnamefont{T.}~\bibnamefont{Nakamura}}
  (\bibinfo{publisher}{World {S}cientific}, \bibinfo{address}{Singapore},
  \bibinfo{year}{1992}), pp. \bibinfo{pages}{244--266}.

\bibitem[{\citenamefont{Danzmann and Rudiger}(2003)}]{Danzmann:2003tv}
\bibinfo{author}{\bibfnamefont{K.}~\bibnamefont{Danzmann}} \bibnamefont{and}
  \bibinfo{author}{\bibfnamefont{A.}~\bibnamefont{Rudiger}},
  \bibinfo{journal}{Class. Quant. Grav.} \textbf{\bibinfo{volume}{20}},
  \bibinfo{pages}{S1} (\bibinfo{year}{2003}).

\bibitem[{\citenamefont{Pretorius}(2005{\natexlab{a}})}]{Pretorius:2005gq}
\bibinfo{author}{\bibfnamefont{F.}~\bibnamefont{Pretorius}},
  \bibinfo{journal}{Phys. Rev. Lett.} \textbf{\bibinfo{volume}{95}},
  \bibinfo{pages}{121101} (\bibinfo{year}{2005}{\natexlab{a}}),
  \eprint{gr-qc/0507014}.

\bibitem[{\citenamefont{Campanelli
  et~al.}(2006{\natexlab{a}})\citenamefont{Campanelli, Lousto, Marronetti, and
  Zlochower}}]{Campanelli:2005dd}
\bibinfo{author}{\bibfnamefont{M.}~\bibnamefont{Campanelli}},
  \bibinfo{author}{\bibfnamefont{C.~O.} \bibnamefont{Lousto}},
  \bibinfo{author}{\bibfnamefont{P.}~\bibnamefont{Marronetti}},
  \bibnamefont{and}
  \bibinfo{author}{\bibfnamefont{Y.}~\bibnamefont{Zlochower}},
  \bibinfo{journal}{Phys. Rev. Lett.} \textbf{\bibinfo{volume}{96}},
  \bibinfo{pages}{111101} (\bibinfo{year}{2006}{\natexlab{a}}),
  \eprint{gr-qc/0511048}.

\bibitem[{\citenamefont{Baker et~al.}(2006{\natexlab{a}})\citenamefont{Baker,
  Centrella, Choi, Koppitz, and van Meter}}]{Baker:2005vv}
\bibinfo{author}{\bibfnamefont{J.~G.} \bibnamefont{Baker}},
  \bibinfo{author}{\bibfnamefont{J.}~\bibnamefont{Centrella}},
  \bibinfo{author}{\bibfnamefont{D.-I.} \bibnamefont{Choi}},
  \bibinfo{author}{\bibfnamefont{M.}~\bibnamefont{Koppitz}}, \bibnamefont{and}
  \bibinfo{author}{\bibfnamefont{J.}~\bibnamefont{van Meter}},
  \bibinfo{journal}{Phys. Rev. Lett.} \textbf{\bibinfo{volume}{96}},
  \bibinfo{pages}{111102} (\bibinfo{year}{2006}{\natexlab{a}}),
  \eprint{gr-qc/0511103}.

\bibitem[{\citenamefont{Campanelli
  et~al.}(2006{\natexlab{b}})\citenamefont{Campanelli, Lousto, and
  Zlochower}}]{Campanelli:2006gf}
\bibinfo{author}{\bibfnamefont{M.}~\bibnamefont{Campanelli}},
  \bibinfo{author}{\bibfnamefont{C.~O.} \bibnamefont{Lousto}},
  \bibnamefont{and}
  \bibinfo{author}{\bibfnamefont{Y.}~\bibnamefont{Zlochower}},
  \bibinfo{journal}{Phys. Rev. D} \textbf{\bibinfo{volume}{73}},
  \bibinfo{pages}{061501(R)} (\bibinfo{year}{2006}{\natexlab{b}}).

\bibitem[{\citenamefont{Baker et~al.}(2006{\natexlab{b}})\citenamefont{Baker,
  Centrella, Choi, Koppitz, and van Meter}}]{Baker:2006yw}
\bibinfo{author}{\bibfnamefont{J.~G.} \bibnamefont{Baker}},
  \bibinfo{author}{\bibfnamefont{J.}~\bibnamefont{Centrella}},
  \bibinfo{author}{\bibfnamefont{D.-I.} \bibnamefont{Choi}},
  \bibinfo{author}{\bibfnamefont{M.}~\bibnamefont{Koppitz}}, \bibnamefont{and}
  \bibinfo{author}{\bibfnamefont{J.}~\bibnamefont{van Meter}},
  \bibinfo{journal}{Phys. Rev. D} \textbf{\bibinfo{volume}{73}},
  \bibinfo{pages}{104002} (\bibinfo{year}{2006}{\natexlab{b}}),
  \eprint{gr-qc/0602026}.

\bibitem[{\citenamefont{Buonanno et~al.}(2006)\citenamefont{Buonanno, Cook, and
  Pretorius}}]{Buonanno:2006}
\bibinfo{author}{\bibfnamefont{A.}~\bibnamefont{Buonanno}},
  \bibinfo{author}{\bibfnamefont{G.}~\bibnamefont{Cook}}, \bibnamefont{and}
  \bibinfo{author}{\bibfnamefont{F.}~\bibnamefont{Pretorius}}
  (\bibinfo{year}{2006}), \eprint{gr-qc/0610122}.

\bibitem[{\citenamefont{Scheel et~al.}(2006)}]{Scheel:2006gg}
\bibinfo{author}{\bibfnamefont{M.~A.} \bibnamefont{Scheel}}
  \bibnamefont{et~al.} (\bibinfo{year}{2006}), \eprint{gr-qc/0607056}.

\bibitem[{\citenamefont{Bruegmann et~al.}(2006)}]{Bruegmann:2006at}
\bibinfo{author}{\bibfnamefont{B.}~\bibnamefont{Bruegmann}}
  \bibnamefont{et~al.} (\bibinfo{year}{2006}), \eprint{gr-qc/0610128}.

\bibitem[{\citenamefont{Caudill et~al.}(2006)\citenamefont{Caudill, Cook,
  Grigsby, and Pfeiffer}}]{Caudill:2006hw}
\bibinfo{author}{\bibfnamefont{M.}~\bibnamefont{Caudill}},
  \bibinfo{author}{\bibfnamefont{G.~B.} \bibnamefont{Cook}},
  \bibinfo{author}{\bibfnamefont{J.~D.} \bibnamefont{Grigsby}},
  \bibnamefont{and} \bibinfo{author}{\bibfnamefont{H.~P.}
  \bibnamefont{Pfeiffer}}, \bibinfo{journal}{Phys. Rev. D}
  \textbf{\bibinfo{volume}{74}}, \bibinfo{pages}{064011}
  (\bibinfo{year}{2006}), \eprint{gr-qc/0605053}.

\bibitem[{\citenamefont{Pfeiffer et~al.}(2003)\citenamefont{Pfeiffer, Kidder,
  Scheel, and Teukolsky}}]{Pfeiffer:2002wt}
\bibinfo{author}{\bibfnamefont{H.~P.} \bibnamefont{Pfeiffer}},
  \bibinfo{author}{\bibfnamefont{L.~E.} \bibnamefont{Kidder}},
  \bibinfo{author}{\bibfnamefont{M.~A.} \bibnamefont{Scheel}},
  \bibnamefont{and} \bibinfo{author}{\bibfnamefont{S.~A.}
  \bibnamefont{Teukolsky}}, \bibinfo{journal}{Comput. Phys. Commun.}
  \textbf{\bibinfo{volume}{152}}, \bibinfo{pages}{253} (\bibinfo{year}{2003}),
  \eprint{gr-qc/0202096}.

\bibitem[{\citenamefont{Brandt and Br{\"u}gmann}(1997)}]{Brandt97b}
\bibinfo{author}{\bibfnamefont{S.}~\bibnamefont{Brandt}} \bibnamefont{and}
  \bibinfo{author}{\bibfnamefont{B.}~\bibnamefont{Br{\"u}gmann}},
  \bibinfo{journal}{Phys. Rev. Lett.} \textbf{\bibinfo{volume}{78}},
  \bibinfo{pages}{3606} (\bibinfo{year}{1997}), \eprint{gr-qc/9703066}.

\bibitem[{\citenamefont{Bowen and York}(1980)}]{Bowen80}
\bibinfo{author}{\bibfnamefont{J.~M.} \bibnamefont{Bowen}} \bibnamefont{and}
  \bibinfo{author}{\bibfnamefont{J.~W.} \bibnamefont{York},
  \bibfnamefont{Jr.}}, \bibinfo{journal}{Phys. Rev. D}
  \textbf{\bibinfo{volume}{21}}, \bibinfo{pages}{2047} (\bibinfo{year}{1980}).

\bibitem[{\citenamefont{Cook}(1994)}]{Cook94}
\bibinfo{author}{\bibfnamefont{G.~B.} \bibnamefont{Cook}},
  \bibinfo{journal}{Phys. Rev. D} \textbf{\bibinfo{volume}{50}},
  \bibinfo{pages}{5025} (\bibinfo{year}{1994}).

\bibitem[{\citenamefont{Pretorius}(2005{\natexlab{b}})}]{Pretorius:2004}
\bibinfo{author}{\bibfnamefont{F.}~\bibnamefont{Pretorius}},
  \bibinfo{journal}{Class.Quant.Grav.} \textbf{\bibinfo{volume}{22}},
  \bibinfo{pages}{425} (\bibinfo{year}{2005}{\natexlab{b}}),
  \eprint{gr-qc/0407110}.

\bibitem[{\citenamefont{Gundlach et~al.}(2005)\citenamefont{Gundlach,
  Martin-Garcia, Calabrese, and Hinder}}]{gundlach:2006}
\bibinfo{author}{\bibfnamefont{C.}~\bibnamefont{Gundlach}},
  \bibinfo{author}{\bibfnamefont{J.}~\bibnamefont{Martin-Garcia}},
  \bibinfo{author}{\bibfnamefont{G.}~\bibnamefont{Calabrese}},
  \bibnamefont{and} \bibinfo{author}{\bibfnamefont{I.}~\bibnamefont{Hinder}},
  \bibinfo{journal}{Class.Quant.Grav.} \textbf{\bibinfo{volume}{22}},
  \bibinfo{pages}{3767} (\bibinfo{year}{2005}), \eprint{gr-qc/0407110}.

\bibitem[{\citenamefont{Nakamura et~al.}(1987)\citenamefont{Nakamura, Oohara,
  and Kojima}}]{Nakamura87}
\bibinfo{author}{\bibfnamefont{T.}~\bibnamefont{Nakamura}},
  \bibinfo{author}{\bibfnamefont{K.}~\bibnamefont{Oohara}}, \bibnamefont{and}
  \bibinfo{author}{\bibfnamefont{Y.}~\bibnamefont{Kojima}},
  \bibinfo{journal}{Prog. Theor. Phys. Suppl.} \textbf{\bibinfo{volume}{90}},
  \bibinfo{pages}{1} (\bibinfo{year}{1987}).

\bibitem[{\citenamefont{Shibata and Nakamura}(1995)}]{Shibata95}
\bibinfo{author}{\bibfnamefont{M.}~\bibnamefont{Shibata}} \bibnamefont{and}
  \bibinfo{author}{\bibfnamefont{T.}~\bibnamefont{Nakamura}},
  \bibinfo{journal}{Phys. Rev. D} \textbf{\bibinfo{volume}{52}},
  \bibinfo{pages}{5428} (\bibinfo{year}{1995}).

\bibitem[{\citenamefont{Baumgarte and Shapiro}(1999)}]{Baumgarte99}
\bibinfo{author}{\bibfnamefont{T.~W.} \bibnamefont{Baumgarte}}
  \bibnamefont{and} \bibinfo{author}{\bibfnamefont{S.~L.}
  \bibnamefont{Shapiro}}, \bibinfo{journal}{Phys. Rev. D}
  \textbf{\bibinfo{volume}{59}}, \bibinfo{pages}{024007}
  (\bibinfo{year}{1999}), \eprint{gr-qc/9810065}.

\bibitem[{\citenamefont{Baker et~al.}(2002{\natexlab{a}})\citenamefont{Baker,
  Campanelli, and Lousto}}]{Baker:2001sf}
\bibinfo{author}{\bibfnamefont{J.~G.} \bibnamefont{Baker}},
  \bibinfo{author}{\bibfnamefont{M.}~\bibnamefont{Campanelli}},
  \bibnamefont{and} \bibinfo{author}{\bibfnamefont{C.~O.}
  \bibnamefont{Lousto}}, \bibinfo{journal}{Phys. Rev.}
  \textbf{\bibinfo{volume}{D65}}, \bibinfo{pages}{044001}
  (\bibinfo{year}{2002}{\natexlab{a}}), \eprint{gr-qc/0104063}.

\bibitem[{\citenamefont{Beetle et~al.}(2005)\citenamefont{Beetle, Bruni, Burko,
  and Nerozzi}}]{Beetle:2004wu}
\bibinfo{author}{\bibfnamefont{C.}~\bibnamefont{Beetle}},
  \bibinfo{author}{\bibfnamefont{M.}~\bibnamefont{Bruni}},
  \bibinfo{author}{\bibfnamefont{L.~M.} \bibnamefont{Burko}}, \bibnamefont{and}
  \bibinfo{author}{\bibfnamefont{A.}~\bibnamefont{Nerozzi}},
  \bibinfo{journal}{Phys. Rev. D} \textbf{\bibinfo{volume}{72}},
  \bibinfo{pages}{024013} (\bibinfo{year}{2005}), \eprint{gr-qc/0407012}.

\bibitem[{\citenamefont{Campanelli
  et~al.}(2006{\natexlab{c}})\citenamefont{Campanelli, Kelly, and
  Lousto}}]{Campanelli:2005ia}
\bibinfo{author}{\bibfnamefont{M.}~\bibnamefont{Campanelli}},
  \bibinfo{author}{\bibfnamefont{B.}~\bibnamefont{Kelly}}, \bibnamefont{and}
  \bibinfo{author}{\bibfnamefont{C.~O.} \bibnamefont{Lousto}},
  \bibinfo{journal}{Phys. Rev. D} \textbf{\bibinfo{volume}{73}},
  \bibinfo{pages}{064005} (\bibinfo{year}{2006}{\natexlab{c}}),
  \eprint{gr-qc/0510122}.

\bibitem[{\citenamefont{Baker et~al.}(2002{\natexlab{b}})\citenamefont{Baker,
  Campanelli, Lousto, and Takahashi}}]{Baker:2002qf}
\bibinfo{author}{\bibfnamefont{J.~G.} \bibnamefont{Baker}},
  \bibinfo{author}{\bibfnamefont{M.}~\bibnamefont{Campanelli}},
  \bibinfo{author}{\bibfnamefont{C.~O.} \bibnamefont{Lousto}},
  \bibnamefont{and}
  \bibinfo{author}{\bibfnamefont{R.}~\bibnamefont{Takahashi}},
  \bibinfo{journal}{Phys. Rev.} \textbf{\bibinfo{volume}{D65}},
  \bibinfo{pages}{124012} (\bibinfo{year}{2002}{\natexlab{b}}),
  \eprint{astro-ph/0202469}.

\bibitem[{\citenamefont{Baker et~al.}(2003)\citenamefont{Baker, Campanelli,
  Lousto, and Takahashi}}]{Baker:2003ds}
\bibinfo{author}{\bibfnamefont{J.}~\bibnamefont{Baker}},
  \bibinfo{author}{\bibfnamefont{M.}~\bibnamefont{Campanelli}},
  \bibinfo{author}{\bibfnamefont{C.~O.} \bibnamefont{Lousto}},
  \bibnamefont{and} \bibinfo{author}{\bibfnamefont{R.}~\bibnamefont{Takahashi}}
  (\bibinfo{year}{2003}), \eprint[http://arXiv.org/abs]{astro-ph/0305287}.

\bibitem[{\citenamefont{Campanelli
  et~al.}(2006{\natexlab{d}})\citenamefont{Campanelli, Lousto, and
  Zlochower}}]{Campanelli:2006uy}
\bibinfo{author}{\bibfnamefont{M.}~\bibnamefont{Campanelli}},
  \bibinfo{author}{\bibfnamefont{C.~O.} \bibnamefont{Lousto}},
  \bibnamefont{and}
  \bibinfo{author}{\bibfnamefont{Y.}~\bibnamefont{Zlochower}},
  \bibinfo{journal}{Phys. Rev. D} \textbf{\bibinfo{volume}{74}},
  \bibinfo{pages}{041501(R)} (\bibinfo{year}{2006}{\natexlab{d}}),
  \eprint{gr-qc/0604012}.

\bibitem[{\citenamefont{Campanelli
  et~al.}(2006{\natexlab{e}})\citenamefont{Campanelli, Lousto, and
  Zlochower}}]{Campanelli:2006fg}
\bibinfo{author}{\bibfnamefont{M.}~\bibnamefont{Campanelli}},
  \bibinfo{author}{\bibfnamefont{C.~O.} \bibnamefont{Lousto}},
  \bibnamefont{and}
  \bibinfo{author}{\bibfnamefont{Y.}~\bibnamefont{Zlochower}},
  \bibinfo{journal}{Phys. Rev. D} \textbf{\bibinfo{volume}{74}},
  \bibinfo{pages}{084023} (\bibinfo{year}{2006}{\natexlab{e}}),
  \eprint{astro-ph/0608275}.

\bibitem[{\citenamefont{Baumgarte et~al.}(2006)\citenamefont{Baumgarte, Brady,
  Creighton, Lehner, Pretorius, and DeVoe}}]{baumgarte:2006}
\bibinfo{author}{\bibfnamefont{T.}~\bibnamefont{Baumgarte}},
  \bibinfo{author}{\bibfnamefont{P.}~\bibnamefont{Brady}},
  \bibinfo{author}{\bibfnamefont{J.}~\bibnamefont{Creighton}},
  \bibinfo{author}{\bibfnamefont{L.}~\bibnamefont{Lehner}},
  \bibinfo{author}{\bibfnamefont{F.}~\bibnamefont{Pretorius}},
  \bibnamefont{and} \bibinfo{author}{\bibfnamefont{R.}~\bibnamefont{DeVoe}}
  (\bibinfo{year}{2006}), \eprint{gr-qc/0612100}.

\bibitem[{NRw()}]{NRwaves}
\bibinfo{note}{NRwaves home page: {\tt https://gravity.psu.edu/wiki\_NRwaves}}.

\end{thebibliography}
\thebibliography{101}

\end{document}